\documentclass{emulateapj}
\usepackage{amsmath,amssymb,amsfonts}
\usepackage{graphicx}
\usepackage{lipsum}
\usepackage{color,soul}
\usepackage{url}

\newcommand{\awg}{\textit{airmass-weighted}}
\newcommand{\ggw}{\textit{galaxy-gw}}

\newcommand{\IUCAA}{Inter-University Centre for Astronomy and 
  Astrophysics, Post Bag 4, Ganeshkhind, Pune 411 007, India}
\newcommand{\WSU}{Department of Physics \& Astronomy, Washington State University,
1245 Webster, Pullman, WA 99164-2814, U.S.A}
\newcommand{\Caltech}{Division of Physics, Math, and Astronomy, California Institute of Technology, Pasadena, CA 91125, USA}

\graphicspath{{./}{figures/}}

\begin{document}

\title{Optimal Optical Search Strategy For Finding Transients in Large Sky Error Regions Under Realistic Constraints}

\author{Javed Rana\altaffilmark{1,\dag}, Shreya Anand\altaffilmark{2}, Sukanta Bose\altaffilmark{1,3}}

\altaffiltext{1}{\IUCAA}
\altaffiltext{\dag}{\email{javed@iucaa.in}}
\altaffiltext{2}{\Caltech}
\altaffiltext{3}{\WSU}

\begin{abstract}

In order to identify the rapidly-fading, optical transient counterparts of gravitational wave (GW) sources, an efficient follow-up strategy is required.  Since most ground-based optical observatories aimed at following-up GW sources have a telescope with a small field-of-view (FOV) as compared to the GW sky error region, we focus on a search strategy that involves dividing the GW patch into “tiles” of the same area as the telescope FOV to strategically image the entire patch.  We present an improvement over the optimal telescope-scheduling algorithm outlined in \cite{ref:rana16}, by combining the tiling and galaxy-targeted search strategies, and factoring the effects of the source airmass and telescope slew, along with setting constraints, into the scheduling algorithm in order to increase the chances of identifying the GW counterpart.  We propose two separate algorithms: the \awg\ algorithm, a specific solution to the Hungarian algorithm that maximizes probability acquired, while minimizing the image airmass, and the slew-optimization algorithm that minimizes the overall slew angle covered between images for the given probability acquired by the optimal telescope-scheduling algorithm in \cite{ref:rana16}.  Using the observatory site of the GROWTH-India telescope as an example, we generate 100s of skymaps to test the performance of our algorithms.  Our results indicate that slew-optimization can reduce the cumulative slew angle in the observing schedule by 100s of degrees, saving several of minutes of observing time without the loss of tiles and probability.  Further, we demonstrate that as compared to the greedy algorithm, the \awg\ algorithm can acquire up to 20 \% more probability and 30 sq. deg. more in areal coverage for skymaps of all sizes and configurations. Our analysis can be straightforwardly extended to optical counterparts of gamma-ray bursts as well as to 
other telescopes or sites.
\end{abstract}


\maketitle

\section{Introduction} \label{sec:intro}

Within the next few years of gravitational-wave (GW) astrophysics, current estimates predict the identification of tens of compact binary coalescence (CBC) sources with electromagnetic (EM) counterparts detectable by ground-based optical telescopes \citep{ref:prospects}.  Specifically, amongst CBC sources, the most promising candidates to contain a visible electromagnetic (EM) counterpart are binary neutron star (BNS) and neutron star-black hole (NSBH) systems.  The recent detection of the binary neutron star system GW170817 using the Hanford-Livingston-Virgo detector network \citep{ref:GW170817_detection} demonstrated that a wealth of information can be gained from combining the electromagnetic and gravitational wave emission from the same source.  For the 2017 detection, GW-EM information was used to identify the source host galaxy \citep{ref:GW170817_detection}, probe properties of the progenitor and remnant \citep{ref:progenitor, ref:remnant}, estimate the Hubble constant \citep{ref:standardsiren}, constrain possible models for the merger and emission \citep{ref:ejecta}, and study the r-process nucleosynthesis processes resulting from the collision of the two neutron stars \citep{ref:smartt17, ref:coughlin18b}.  The 2017 BNS detection also marked the first confirmed detection of a kilonova, the bright, infrared or UV emission hypothesized to result from the r-process nucleosynthesis occurring during a BNS or NSBH merger \citep{ref:GW170817_MMA}.  This study focuses on the problem of identifying optical counterparts to GW or GRBs localized to large sky-error regions.

Due to the proximity of the binary neutron star system to Earth, the gravitational-wave source was localized to a sky area and volume of 28 sq. deg. and 380 Mpc$^3$ respectively, making it the most well-localized amongst all past gravitational wave detections \citep{ref:GW170817_detection}.  The fact that the source was localized to the nulls of Virgo's antenna pattern significantly reduced the localization area from an ordinary two-detector localization \citep{ref:GW170817_detection}.  IM2H, the observing team that first detected the optical counterpart to GW170817 about 10 hours after the gravitational wave detection, employed a strategy of targeting known galaxies within the source's localization volume \citep{ref:sss17, ref:GW170817_MMA}.  Though we anticipate that some future BNS detections will be as nearby and as well-localized as GW170817, not all the GW detections of BNS systems will have a bright optical counterpart.  In fact, only a small fraction of sources will be optimally oriented, maximizing the GW amplitude and favoring the detection of a coincident on-axis afterglow (\cite{ref:metzgerberger12, ref:petrillo13, ref:barneskasen13, ref:kasen15, ref:metzger17}). Thus, we do not anticipate the search for optical counterparts to BNS sources to be as straightforward as it was during the GW170817 detection.

The approach for detecting counterparts to GW events differs based on the wavelength of the emission, due to the fact that each type of emission is observable for a different duration of time.  Optical counterparts to BNS and NSBH sources, which could either be kilonovae \citep{ref:metzger17}, or optical afterglows \citep{ref:ascenzi18, ref:ghosh&bose13}, fade rapidly, and could last anywhere from hours to days after the gravitational wave source is detected \citep{ref:metzgerberger12}.  Using optical telescopes to observe the source as early as possible after the GW detection will maximize the information gained from the source spectrum.  During the upcoming third GW observing run, the two advanced LIGO \citep{ref:aLIGO} interferometers and the advanced Virgo \citep{ref:aVirgo} interferometer will be ``online" and taking data for one year.  Based on the expected number and sensitivity of gravitational wave detectors detecting BNS and NSBH sources in this Advanced Detector Era (ADE), GW localization sky error regions are expected to span a few tens to a few hundreds of square degrees \citep{ref:nissanke13, ref:singer14}, while most wide-field optical observatories following up GW sources tend to have fields-of-view of less than a degree to a few square degrees, requiring an extensive search within the sky localization region in order to locate the optical transient \citep{ref:singer14, ref:cornishlittenberg15, ref:essick15, ref:klimenko16, ref:rana16}. Modern optical surveys with large FOV telescopes such as ZTF \citep{ref:bellm14}, ATLAS \citep{ref:ATLAS}, Pan-STARRS \citep{ref:Pan-STARRS1}, and LSST \citep{ref:LSST} are considered ideal in this scenario; however, implementing optimized search strategies for smaller-FOV optical telescopes will increase the odds of identifying a coincident optical counterpart to a given BNS/NSBH trigger by strengthening the overall telescope follow-up network.  In the remainder of the paper, our discussion of ``optical telescopes" will primarily concern telescopes for which a scheduling strategy is most relevant (i.e. FOV $<$ 10 deg$^2$), though the same methods could be applied to optimize the scheduling of larger-FOV telescopes.  

Optical telescopes usually employ either the galaxy-targeted or the tiling strategies to search for GW counterparts.  The former strategy requires identification of all of the galaxies within the sky error region (and therefore, a near-complete galaxy catalogue in the vicinity of the source), and determines the probability of the galaxy containing the GW trigger, based on its mass or luminosity.  Then, ranking the galaxies in order of their properties, the algorithm will provide the telescope with the coordinates of each of the galaxies within the region and the order in which the telescope should point at them \citep{ref:gehrels16}.  

The tiling strategy has four main steps: placing tiles, allocating time, scheduling, and evaluating efficiency \citep{ref:coughlin18b}.  Amongst different tiling methods described in \cite{ref:coughlin18b}, the ranked tiling method \citep{ref:ghosh16} that we employ in this paper involves dividing the GW sky error region into ``tiles" that are the size and shape of the field-of-view of the observing telescope.  After the placement of tiles, the next step is to perform time allocation.  \cite{ref:salafia17} and \cite{ref:chan17} propose time allocation based on counterpart lightcurve models, while \cite{ref:coughlin&stubbs16} derive scaling relations for time allocation based on GW likelihood and galactic extinction.  The two algorithms we present in this work use two different methods of allocating exposure time to tiles. Once exposure time has been allocated to each tile, one can schedule the tiles for observation.  Scheduling involves running an optimization algorithm on the tiles to maximize probability and patch area coverage, and generating a list of the coordinates of the observable tiles and the order in which the telescope should observe them \citep{ref:rana16}.  Finally, one can evaluate the process by running simulations and determining the amount of GW probability acquired by tiling the patch.  Our optimization algorithms combine the tiling and the galaxy-targeted strategies.

Past work on optimizing telescope scheduling \citep{ref:rana16}, uses the tiling method for the purpose of optical follow-up of GW transients, and describes three main algorithms for ranking tiles within the GW sky error region in order to generate an observing sequence.  They are outlined in brief, below:

\begin{itemize}
\item greedy algorithm - ranks tiles in descending order according to probability; doesn't account for setting tiles
\item setting algorithm - selects the highest probability tiles in each setting window, starting from the earliest setting tiles
\item optimal algorithm - reorders the observing sequence generated by the setting algorithm to observe the highest probability tiles first, without losing setting tiles
\end{itemize}

\cite{ref:rana16} demonstrates that while the optimal and setting array algorithms have equivalent performance in terms of probability and area coverage within the patch, the optimal array provides an advantage over setting array as it schedules higher probability tiles to be observed first.  This work builds on the work of \cite{ref:rana16} by considering the effects of slew and airmass on the existing algorithms, and optimizing over both parameters.  

Slew is the process of rotating a telescope to observe different regions of the sky.  The slew angle between two tile observations is equivalent to the solid angle between two points on a spherical surface, simply given by the spherical law of cosines as:
\begin{equation} \sigma_{slew} = cos^{-1}(\sin alt_1 * \sin alt_2 + \cos alt_1 * \cos alt_2 * \cos \Delta az) \end{equation}

\noindent
Here, alt$_1$ and alt$_2$ and $\Delta \rm az = |az_2 - az_1|$, correspond to the two altitude coordinates and the difference between the azimuthal coordinates of each of the points on the sky, assuming a celestial sphere Earth-based coordinate system.  The previous optimization algorithms neither take into account the effect of telescope slewing within the observing sequence nor factor in the time spent slewing between individual tile exposures.  Thus, in this paper, we demonstrate the effect of slewing in the existing algorithms, and propose an alternate algorithm that minimizes the slew between tile observations without loss of tiles or probability.    

We also calculate and compare the cumulative airmass amongst different algorithms for various patch-observatory configurations.  Airmass is the path length for light from a celestial source to pass through the atmosphere; near the horizon, where light is attenuated by scattering and absorption, the airmass is at its maximum, while at the observatory's zenith, it is unity. However, many of the tiles scheduled to be observed with high airmass will require a long exposure time in order to resolve the source, while tiles closer to the zenith could be observed within a brief exposure time.  Thus we implement a modification to the the optimal algorithm that will minimize the airmass at which tiles are observed and adjust each tile exposure time based on tile airmass.  

In addition to the algorithms presented in \cite{ref:rana16}, we discuss the following algorithms in this paper:

\begin{itemize}
    \item modified optimal algorithm - this algorithm is a modification of the optimal algorithm mentioned earlier that factors in the slew time into the observing schedule
    \item slew-optimization algorithm - we propose a new algorithm to optimize over slew and setting constraints to acquire the maximum probability in the patch
    \item \awg\ algorithm - we propose another new algorithm to optimize over the airmass and setting constraints to acquire the maximum \awg\ probability in the patch
\end{itemize}

Like in the previous algorithms, all of the algorithms discussed in this paper account for ground-based visibility constraints and only schedule tiles that are above the horizon.  We run simulations at the locations of the optical observatories GROWTH-India \citep{ref:GROWTH, ref:GROWTH-India} and ZTF to systematically compare the performance of the slew optimization and the \awg\ algorithm with that of the modified optimal and greedy algorithms.

\section{Description of Problem} \label{sec:description}

While previous methods operated under the assumption that the slew time was accounted for in the tile exposure, our calculations demonstrate that the time the telescope spends slewing can cut significantly into the time available for observing tiles in the greedy, setting, or optimal sequence, preventing the telescope from covering all of the tiles provided by the optimal array.  Depending on the shape of the patch, the telescope FOV, and the slew rate, the telescope can spend anywhere from 10s of minutes to over an hour in slewing to tile patches of a few hundreds of square degrees.  In general, this problem of slew times exceeding a tile exposure resulting in loss of tiles is relevant for telescopes with camera readout times that are less than the average slew time, and telescope-patch configurations that require several telescope pointings in order to cover the 95 percent credible region.  For the rapidly fading transients that optical telescopes seek to follow up, the loss of even a few minutes could make the difference between detection and non-detection of the EM counterpart of a GW source \citep{ref:rana16}.

\begin{figure*}[h!]
\plottwo{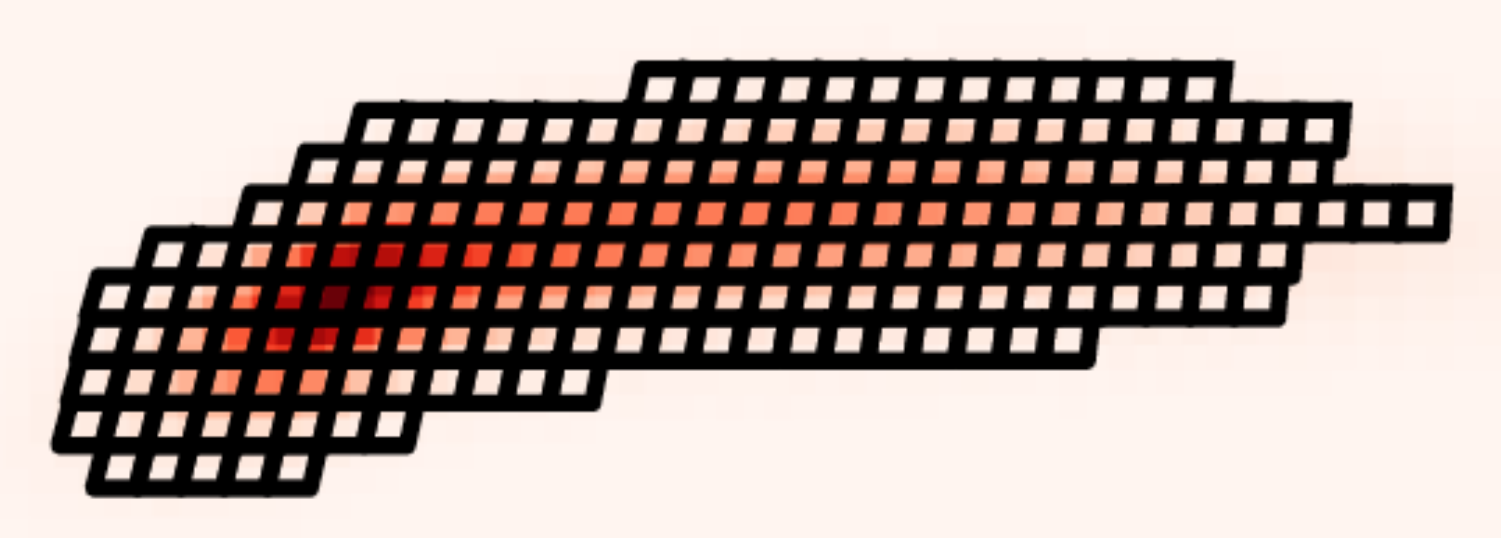}{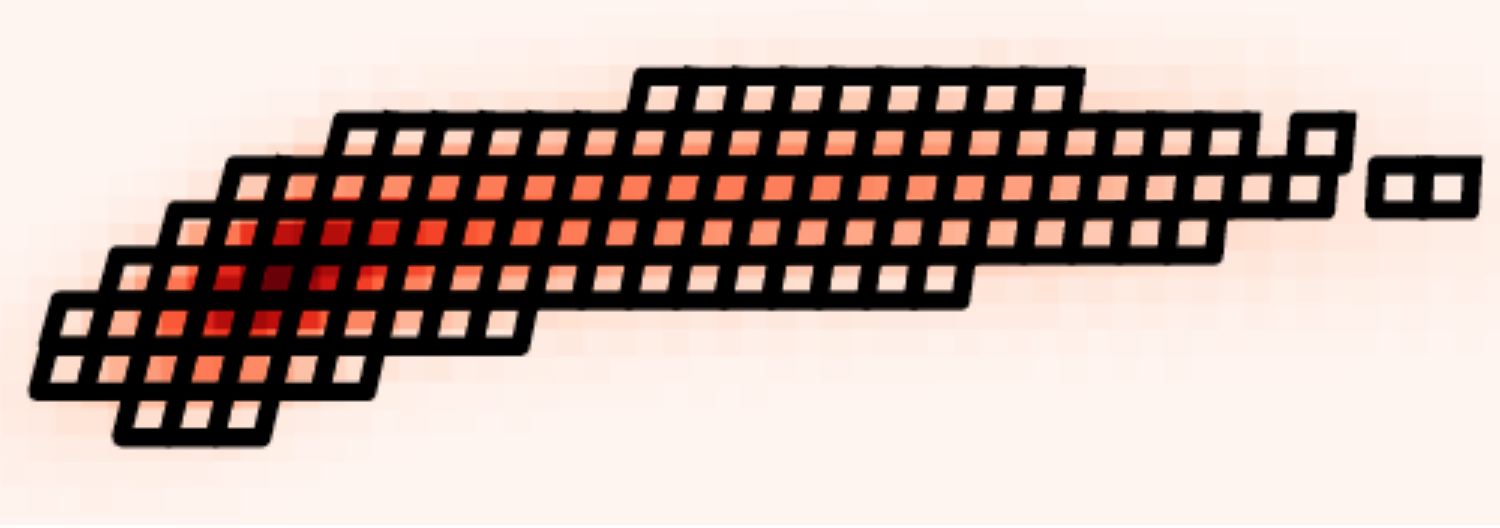}
\caption{Tile coverage on a 94 deg$^2$ sky patch with a 0.5 deg$^2$ FOV at the location of the GROWTH-India telescope (Ladakh, IN) using the optimal algorithm, without accounting for slew (blue curve) and accounting for slew (orange curve).  We assume a tile exposure time of 300 s, based on the current estimates for the exposure time needed to observe a 19th mag source (S. Srivastav).  The telescope has a slew rate of 2 deg/s.  Each tile, indicating an image taken by the telescope, is represented using a black square; untiled regions remain reddish, the color of the patch.  The original algorithm schedules the sample observatory to 186 of the 197 total tiles; by accounting for slew, the optimal schedule can only cover 114, a loss of 62 tiles from the original 186 it was scheduled to cover.  As is demonstrated above, in this telescope-patch configuration, the original optimal algorithm \citep{ref:rana16} which does not account for slewing, overestimates the telescope's ability to tile the patch.}
\label{fig:coverage_optimal}
\end{figure*}

\begin{figure}[h!]
\includegraphics[width=\columnwidth]{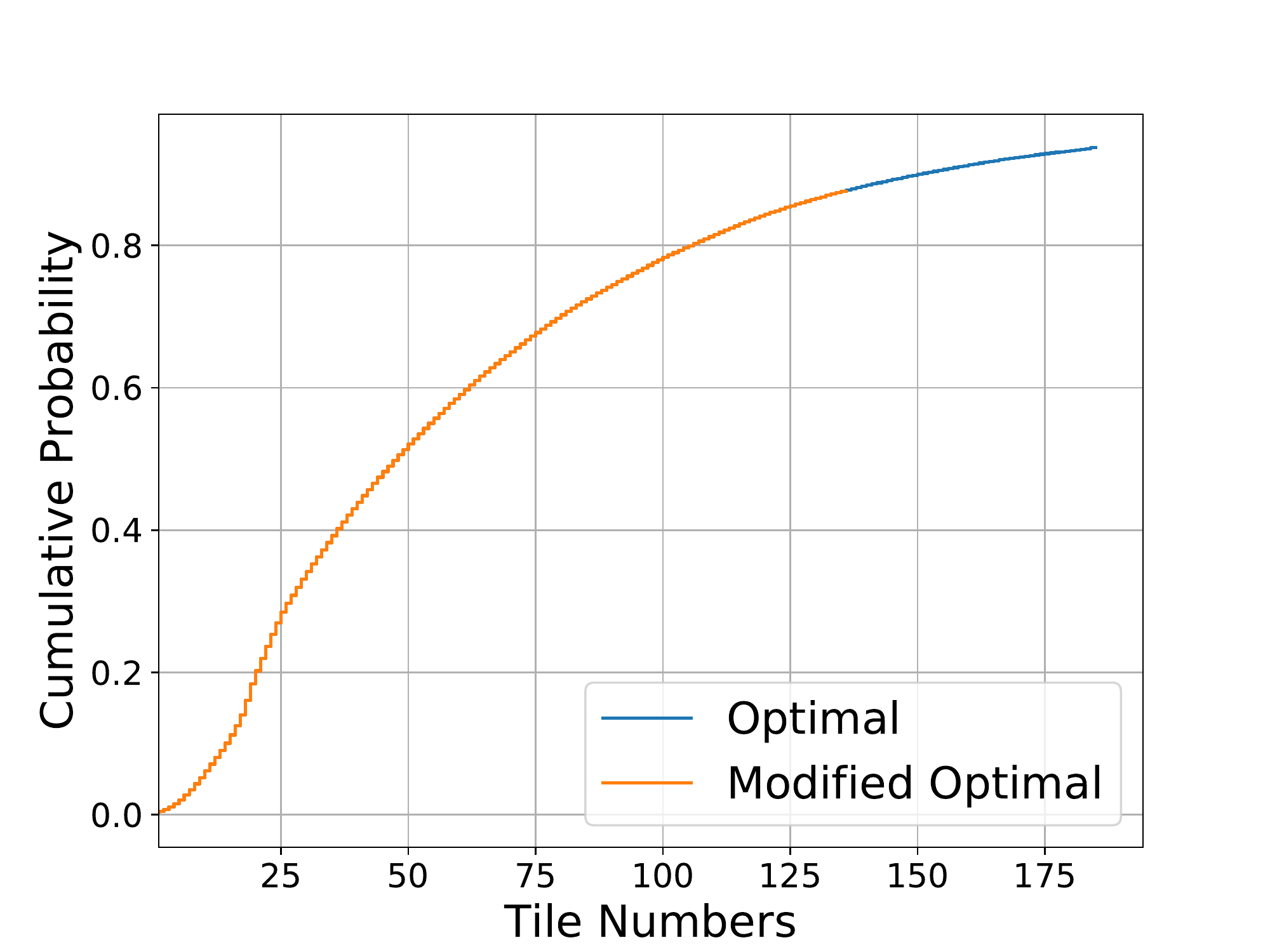}
\caption{Cumulative probability on a 94 deg$^2$ sky patch using the optimal algorithm, without accounting for slew (blue curve) and the modified optimal algorithm (orange curve).  The telescope has a slew rate of 2 deg/s. The plots chart the probability covered within the patch as a function of the number of images taken.  The original algorithm predicts that the sample observatory can cover a probability of 93.7\%; accounting for slew, the observatory is only able to cover 82.1\% probability.  As is demonstrated above, the amount of probability acquired for this telescope-patch configuration after accounting for slew is much less than predicted by the original optimal algorithm \citep{ref:rana16}.}
\label{fig:cumprob_optimal}
\end{figure}

In order to determine how tiles lost to slewing affect the overall performance of our existing algorithms, we run the optimal algorithm using the FOV, location, and other properties of the GROWTH-India telescope (see Table 1) to tile a 94 square degree patch.  First we run the algorithm without accounting for the slew time, and the second time, the algorithm calculates the overall time spent slewing, and removes tiles appropriately from the observing sequence.  We refer to the optimal algorithm with slew accounted for as the modified optimal algorithm.  Neglecting to account for slew time within the telescope scheduling could negatively affect the overall algorithm performance. If the total time spent slewing exceeds a single or several tile exposures, one or more tiles could be scheduled after they have already set.  When we account for slewing, we omit these already-set tiles; as a result, accounting for slew demonstrates that the total probability acquired at the end of the observation is, in reality, \textit{lower} than predicted by the original code.  As demonstrated by Figure \ref{fig:coverage_optimal}, in certain cases, the original optimal algorithm grossly overestimates the telescope's ability to tile the patch when it does not account for time lost due to slewing.  The comparison plots of the optimal algorithm tile coverage and cumulative probability are shown in Figures \ref{fig:coverage_optimal} and \ref{fig:cumprob_optimal}.

\begin{table*}[h]
\begin{center}
\begin{tabular}{cccccc}
\hline
\hline
Telescope Name & Location & FOV size & latitude & longitude & Altitude \\
\hline
 & & (deg$^2$) & & & (m) \\

 \hline
 GROWTH-India & Ladakh, IN & 0.49 & 32.78 & 78.96 & 4500 \\
 \hline
 Himalayan Chandra Telescope & Ladakh IN & 0.25 & 32.78 & 78.96 & 4500 \\
 \hline
 Swope & Atacama Region, CL & 3.68 & -29.02 & -70.67 & 2380 \\
 \hline
PTF & Mt. Palomar, CA & 7.80 & 33.35 & -116.86 & 1712 \\
 \hline
 \hline
 \end{tabular}
 \caption{Properties and parameters of the optical observatories used in simulations to test the slew optimization algorithm.}
 \end{center}
\label{table:observatories}
\end{table*}

One of the limiting factors in being able to identify potential transient counterparts to gravitational wave events is the ability to resolve faint and distant sources.  Depending on the size of the telescope field-of-view, within a given tile, there could be multiple galaxies, each with varying distance to Earth. In order to conduct a less-biased search, we choose to adopt a strategy such that the faintest known galaxies within the tile are resolvable.  Furthermore, an optical survey is likely to miss even bright sources within tiles that are about to set, as the effects of atmospheric extinction attenuate the light from the source, making it more difficult to observe. For these reasons, when generally conducting optical observations, each target imaged is allotted a different exposure time.  Therefore, the tile exposure time, though previously uniformly assigned to all tiles, must change depending on the most distant galaxies within the tile and the tile airmass.  

\section{Methods} \label{sec:methods}

\subsection{Slew Optimization} \label{sec:slewopt}

Past work on slew optimization involved the radio follow-up of blazars \citep{ref:maxmoerbeck14}, as the several hours large radio interferometers can spend slewing cuts significantly into the overall source observing time.  In the case of radio follow-up, because the main constraints imposed are the rising and setting times of the patch, optimizing the slew involved applying a slight modification to the well-known ``traveling salesman'' algorithm to minimize the slew path between the tiles.  However, as additional factors such as the telescope FOV, sunrise, and sunset are pertinent for optical follow-up, we cannot simply apply the traveling salesman algorithm on the entire patch to minimize the slew. Instead, we characterize the patches based on the patch visibility, determine whether slew optimization will be possible, and then apply the algorithm on them, with the setting and rising constraints imposed.  In general, slew optimization will only be possible for cases in which all tiles in a given patch rise above the telescope's horizon.  As with the optimal array algorithm \citep{ref:rana16}, our starting point is the setting array algorithm.  We sketch the process of slew optimization for optical follow-up of GW transients as follows:

\begin{enumerate}

\item Convolve a galaxy catalogue with the GW probability distribution on the patch. We modify the tile probability based on the total mass of galaxies contained within each tile, eliminating tiles from the patch that do not contain any galaxies.  This step is optional for the slew-optimized algorithm.

\item Run the setting array algorithm \citep{ref:rana16} on the patch. The setting array algorithm provides the set of maximum probability tiles within the available observation time.

\item Calculate the total slew angle and airmass of the tile array.  We assign the variable t$_{\rm expo}$, a single tile exposure time, such that it will include the slew time of any tiles under 20 degrees when initially selecting tiles for the optimal sequence.  The algorithm operates under the assumption that all slews less than 20 degrees will take the same amount of time as a 20 degree slew, as it is difficult for telescopes to maintain exact slew times for small slew angles ($\sigma_{\rm slew} \leq 20$), due to the acceleration and deceleration of the telescope. Therefore we allot a time gap such that $\rm t_{\rm gap} = \rm t_{\rm expo} + \rm T_{20}$, where T$_{20}$ is the approximate time taken for slews less than 20 degrees.  Here, $\rm T_{20} = (20 \deg) / (v_{\rm slew})$, where $\rm v_{\rm slew}$ is the telescope slew rate.

\item Calculate amount of gained time and use it to acquire additional probability. For cases in which slew optimization is possible, after running the traveling salesman algorithm on the patch, we apply the time gained from optimization to search for additional unobserved tiles.  We define the gained time as follows: 

\begin{equation} 
\begin{split}
t_{gained} = (20 / v_{slew}) * N_{tiles} - t_{slew, exact} + \\ (({T > 20})_{optimal} - ({T > 20})_{slew})
\end{split}
\end{equation}

where the first term is the difference between the accounted-for slew time and the exact amount of time spent slewing, and the second term records difference between time spent performing slews greater than 20 degrees, before and after slew-optimization.  

\item Re-shift the tiles by their exact slew times.  Because the slew optimization algorithm makes use of the gained time to observe additional visible tiles when possible, we space the tile observation times by their exact slew times, accounting for the time spent slewing between successive observations within the schedule itself.  Both our modified optimal and slew-optimization algorithms space tile observations by the slew times between tiles such that the overall slewing time in the final schedule does not exceed a single tile exposure time.  

\item Remove already-set tiles.  The loss of total coverage and cumulative probability after accounting for slew arises solely from the fact that time-shifting the tile observations by their slew time results in a number of tiles being scheduled after they have already set.  Thus we only include tiles in the schedule where the observation time is earlier than the tile set time.  For tiles scheduled before they have risen, we shift their observation times by one tile exposure.  This process is performed iteratively throughout the optimization.  Because we have included the slewing time within the observation time, some tiles that could previously be observed using the original optimal algorithm will be lost due to setting.

\end{enumerate}

\subsection{Airmass and Exposure Time} \label{sec:airmass}

Our objective is to maximize the total probability coverage based on the setting time and the airmass of the tiles on the GW localization. The tiles on the localization move as the Earth rotates, and the airmass of the tiles change as the altitude of the tiles change with time. The airmass of a given tile changes at each observation time. Therefore, we propose an algorithm to maximize the total probability coverage over the setting and airmass of the GW localization. We use the Hungarian optimization algorithm~\citep{Munkres1957} to get the optimal solution for this problem. In our algorithm, we define a new probability called the \awg\ probability. The \awg\ probability of a tile is the ratio between the \ggw\ probability and the airmass of that tile. Although the \ggw\ convolved (or GW) probability is fixed at all times for each tile in the localization, the \awg\ probability of each tile varies with time. We determine the \awg\ probability for each tile at all different times, using the time-dependent airmass, and use our algorithm to select the schedule of tiles that will maximize the \awg\ probability. We demonstrate the algorithm point wise below:

\begin{enumerate}

\item Convolve the GW probability with a galaxy catalogue.  We modify the GW probability based on the total mass of galaxies contained in the given localization. We make a grid of tiles based on the convolved probability, eliminating those tiles that do not contain any galaxies. This is an optional step in our algorithm. 

\item Account for airmass.  The airmass is given by 
\[
am(t) =
\begin{cases} 
 \frac{1}{cos(90 - alt(t))}, & \text{if } alt(t) \geq alt(horizon) \\
 \infty,                     & \text{if } alt(t)<alt(horizon) \label{eqn:airmass}
\end{cases}
\]

where $am(t)$ is the airmass as a function of time and $alt(t)$ is the altitude in degrees as a function of time and $alt(horizon)$ is the altitude of the observing horizon of the telescope. The \awg\ probability of a tile is inversely proportional to its airmass:
\begin{equation} p_{amw} = Norm\left[ \frac{p_{gw}}{am} \right] \label{eqn:airmass_prob} \end{equation}
where $p_{gw}$ is the GW (or \ggw) probability, $p_{amw}$ is the \awg\ probability, and $Norm$ is the normalization constant chosen such that the total airmass-weighted probability is unity.  Our strategy here is to devote less telescope time to tiles at higher airmass, so that low-airmass high-probability tiles are prioritized for observation. At any given time, if the altitude of a tile becomes less than the observing horizon, we make the airmass value infinity so that the \awg\ probability will become zero. Using Eqns. \ref{eqn:airmass} and \ref{eqn:airmass_prob}, we account for the setting and rising of all the tiles in the localization.

To illustrate how the tile airmass affects p$_{amw}$, we can consider an example where two tiles with the same \ggw\ probability are at different altitudes. If one tile is at an altitude of 30 degrees from the horizon and the other is located at the zenith, the first tile will have double the airmass of the tile at the zenith, so the \awg\ probability of that tile will be half of the \awg\ probability of the tile at the zenith.

We calculate $p_{amw}$ for every tile for all time steps from the observation start time to the observation end time. We make a table, where the rows represent the time steps (one time step is equivalent to one exposure) and the columns represent the tiles in the localization. One element in the $ith$ row and $jth$ column is the \awg\ probability of finding the source at the $jth$ tile at the $ith$ time step.

\item Use the Hungarian algorithm to maximize the \awg\ probability. If there exists more than one optimal solution, we choose the solution where higher probability tiles with lower airmass are scheduled earlier.  

\item Allocate exposure time. For a galaxy targeted hunt, we adjust each tile exposure time based on the most distant galaxy within that tile, as the source's flux decreases as its distance squared:
\begin{equation} t_{exp} = \frac{d_{gal}^2}{d_{0}^2} t_{exp0} \end{equation}
where $t_{exp}$ is the adjusted exposure time of the tile, $t_{exp0}$ is the starting exposure time assigned to all tiles,  d$_{0}$ is the maximum distance within the 3D GW localization within the tile, and d$_{gal}$ is the distance to the farthest galaxy in the tile. If the telescope FOV is large enough, all tiles will contain both nearby and distant galaxies such that the required exposure time to resolve distant galaxies in each tile will average out.

Then, we apply a second adjustment to the tile exposure time based on its airmass:
\begin{equation} t_{exp} = am(t) t_{exp0} \end{equation}
where am(t) is the tile airmass, and t$_{exp0}$ could either be the starting exposure time or the galaxy-adjusted exposure time for the tile, depending on whether the galaxy convolution step was performed.  In allocating our exposure time, we do not account for the fading counterpart lightcurve.

The starting exposure time is a user-defined input into our code. One should set the starting exposure time by first selecting one tile in the localization, and then determining the amount of exposure time needed in order to observe a source at some desired magnitude out to the maximum distance within the tile (based on the 3D GW localization). Our code assigns that input exposure time as the initial exposure time for all tiles, and then applies the adjustments described above.  In our simulation, we use the highest probability tile within the localization to set our starting exposure time.

\end{enumerate}

Combining the \awg\ method with the slew optimization method is very complicated, as the \awg\ method breaks the order determined by the slew optimization algorithm that will minimize the slew between tiles. For this reason, we do not attempt to combine the slew and airmass optimizations.  Instead, we show our results for the \awg\ algorithm separately from the slew optimization method, for which we only adjust the exposure times of the tiles appropriately in the finalized sequence.  In this section, the two new methods we have presented are to be considered as alternative methods by which to schedule observations.  

The question of which algorithm to use in a given situation now becomes relevant. The most direct way to determine which scheduling method is most suitable for a given telescope-patch configuration is to run both algorithms on the patch, and choose the algorithm that covers the most GW probability. On a standard 2.4 GHz processor, the slew-optimization algorithm takes $\sim$ 3-5 minutes to complete and the \awg\ method takes less than a minute to complete, so running both algorithms before the start of the observation does not present any significant overheads in time. The cumulative GW probability acquired is directly comparable between both algorithms, as the GW probability relies only on the GW skymap to inform about the true location of the source.

For the galaxy targeted search, an additional complication is the incompleteness of galaxy catalogs such as CLU and Glade \citep{ref:CLU, ref:glade} out to distances larger than $\sim$ 200 Mpc.  However since most optical telescopes are only sensitive out to about 200 Mpc, this should not significantly bias the results.

\section{Results} \label{sec:results}

In this section, we use both case studies and simulations to test the performance of our two algorithms.  For the purpose of this study, the telescopes we select (GROWTH-India, HCT, Swope, and PTF) are wide-field optical telescopes that have small to medium FOV sizes and are apt for rapid follow-up of transient events.  Table 1 shows the locations and parameters of each of the telescopes used.

\begin{figure}[ht]
\includegraphics[width=\columnwidth]{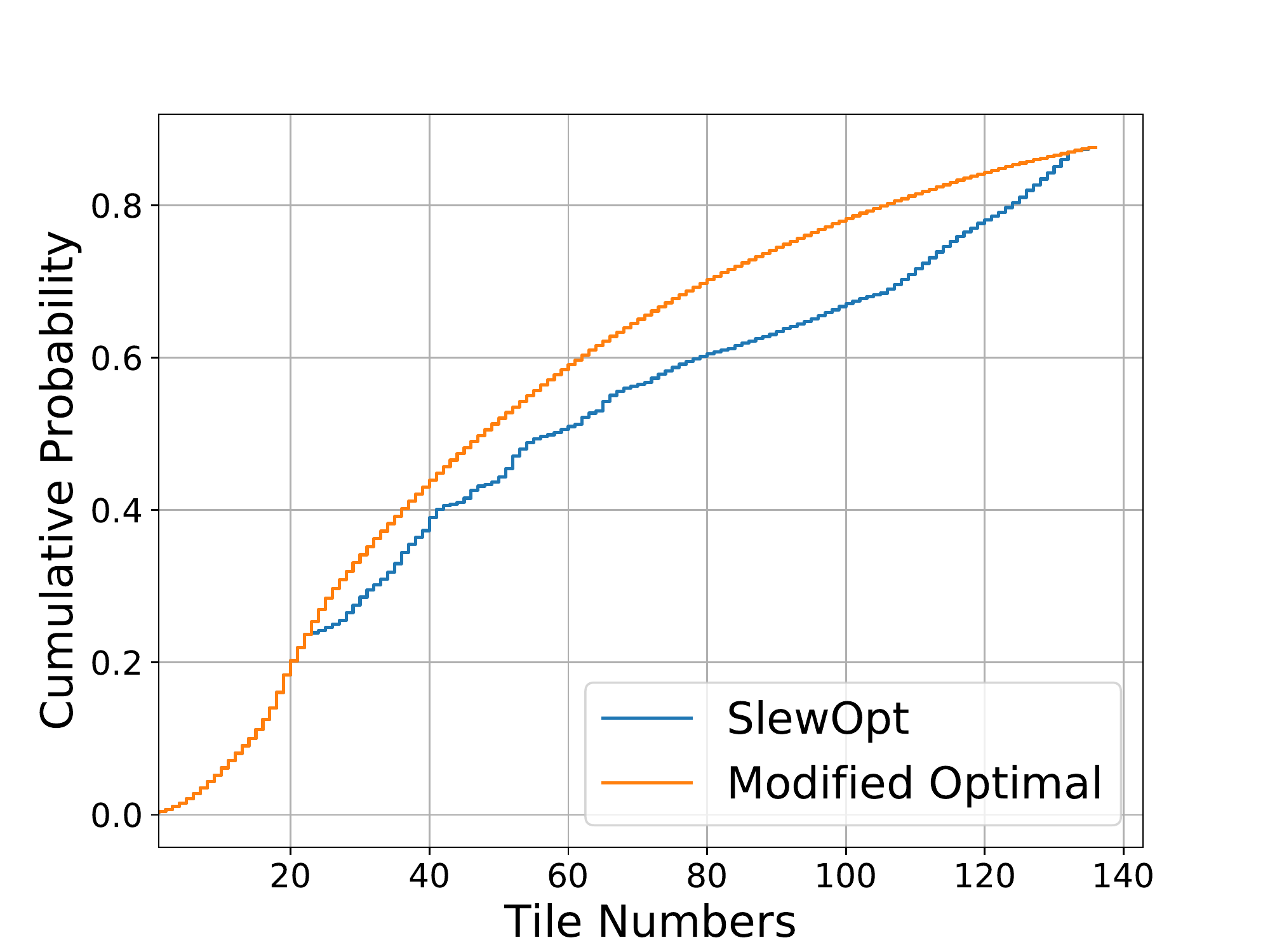}
\hspace{0.5mm}
\caption{Cumulative probability on a 94 deg$^2$ sky patch using the modified optimal algorithm that accounting for slew (left) and slew optimization algorithm (right).  The telescope has a slew rate of 2 deg/s. The plots chart the probability covered within the patch as a function of the number of images taken. The slew-optimization algorithm acquires about the same amount of probability as the modified optimal algorithm (See Figure \ref{fig:cumprob_optimal})}.
\label{fig:cumprob_optimal_slew}
\end{figure}

\begin{figure}[htb]
\label{fig:cumslew_optimal_slew}
\includegraphics[width=\columnwidth]{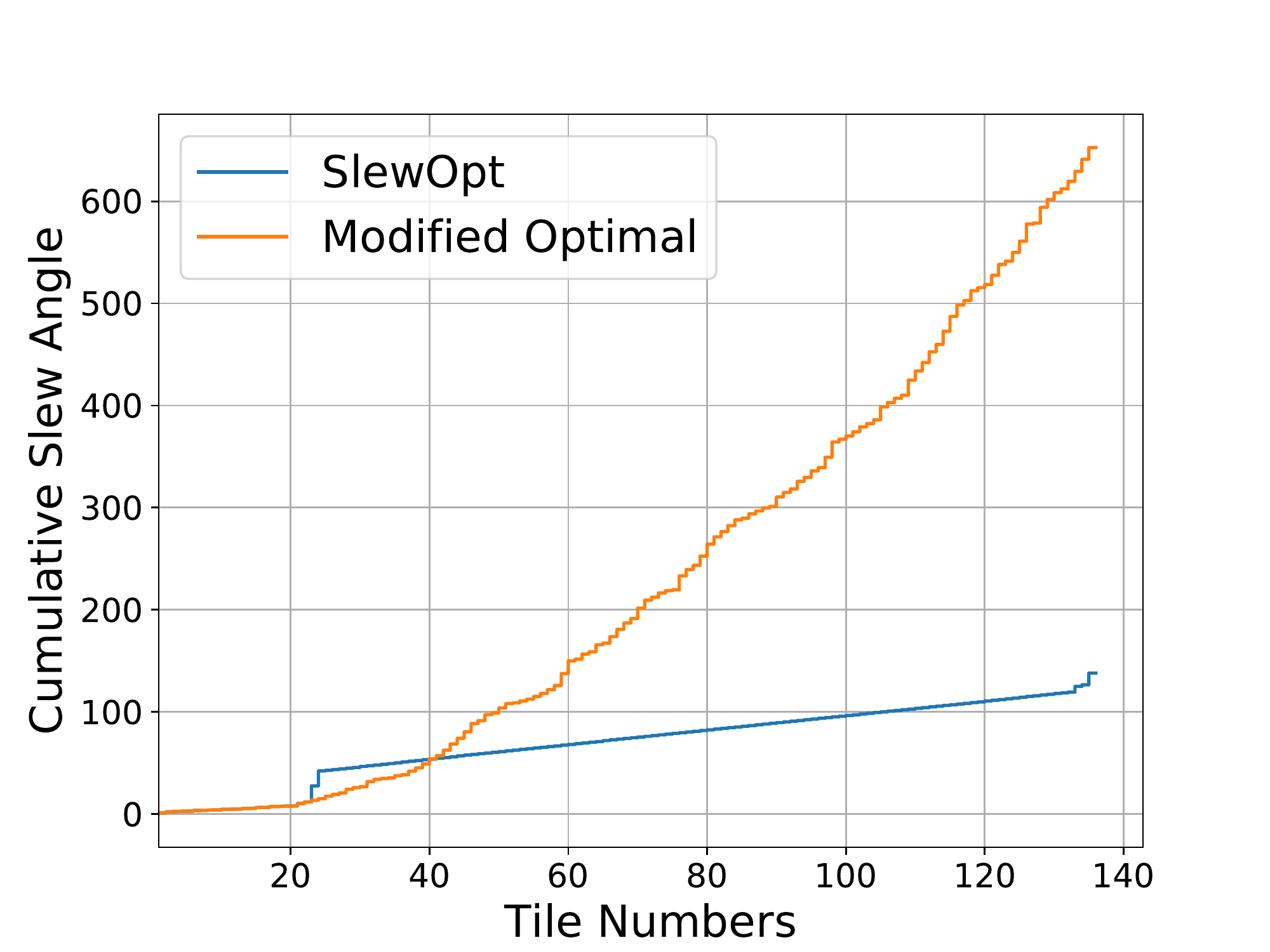}
\hspace{0.5mm}
\caption{Cumulative slew angle on a 94 deg$^2$ sky patch using the optimal algorithm, accounting for slew (orange curve) and slew optimization algorithm (blue curve).  The telescope has a slew rate of 2 deg/s. The plots chart the telescope's cumulative slew angle over its observation of the patch.  Around tiles 25 and 125, there are larger slew jumps.  After optimizing for slew, the telescope's cumulative slew angle reduces by about 300 degrees, allowing additional time for further observation.}
\end{figure}

We revisit the example referred to earlier in Sec. \ref{sec:description} to see how the slew-optimization algorithm performance differs from that of the modified optimal algorithm for the GROWTH-India telescope.  We can compare the cumulative probability in each case using Figure \ref{fig:cumprob_optimal_slew}.  Without taking slewing time into account the telescope is capable of covering all of the tiles in the patch, covering a cumulative probability of 93.7\% (see Fig. \ref{fig:coverage_optimal}).  On the other hand, the modified optimal algorithm can only cover a probability of 82.1\%.  The slew optimization algorithm is designed such that it cannot perform worse than the modified optimal algorithm, in terms of probability and coverage; for the cases in which the slew cannot be further optimized, the algorithm will return the same tile schedule as the optimal algorithm.  In this case, the slew-optimization algorithm acquires about the same amount of probability (82.1\%) as the modified optimal algorithm, though both algorithms far surpass the amount of probability covered by the greedy algorithm.  The key advantage of the slew-optimized algorithm is demonstrated in Figure \ref{fig:cumslew_optimal_slew} by the cumulative slew angle being reduced from above 600 degrees to about 140 degrees as a result of slew-optimization.  The time saved in slewing aids in more rapid observation of the patch, which is important especially when there are multiple targets to observe in one night.

By comparing cumulative slew angles between the slew-optimized array and the modified optimal array, we can see that the curve shapes differ quite drastically.  The optimal method schedules tiles to be observed with small slew jumps between observations, resulting in a relatively smooth increase in overall slew angle with tile number.  On the other hand, the slew-optimized method is mostly linear with two jumps around the 20th and 130th tile in the schedule.  In effect, the slew-optimization method attempts to schedule as many tiles consecutively as possible before jumping to a different region of the patch. Figure \ref{fig:cumslew_optimal_slew} displays the behavior of the slew optimization algorithm as expected.

\begin{figure}[h!]
\includegraphics[width=0.8\columnwidth, angle=270]{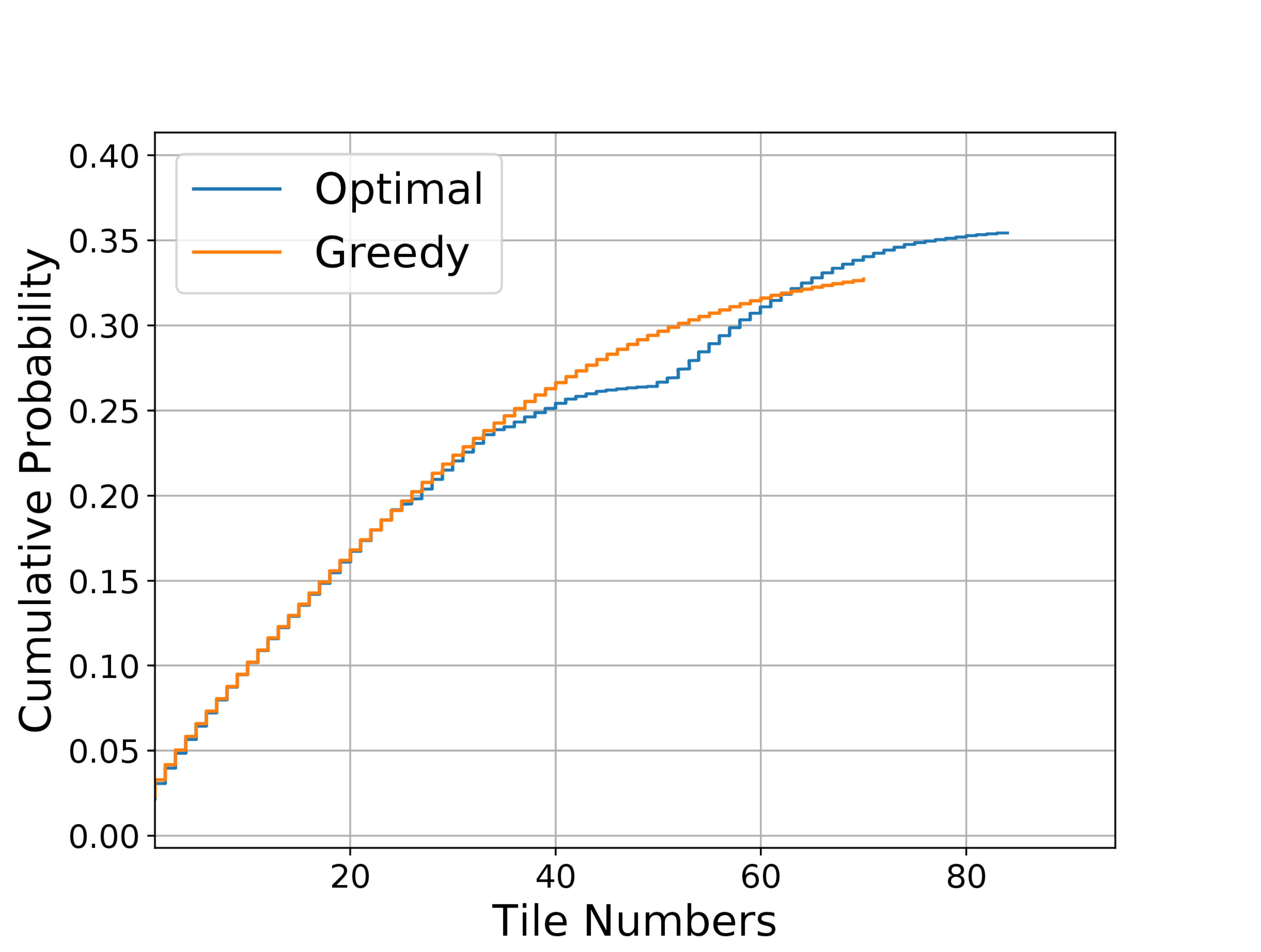}
\caption{Cumulative probability as a function of tile number at the location of the PTF observatory using the \awg\ algorithm. The plots chart the probability covered within the patch as a function of the number of images taken. The \awg algorithm (blue curve) covers a probability of 36.5\%, while the greedy algorithm (orange curve) covers a probability of 32.8\%. When comparing the performance of this modified optimal algorithm with the greedy algorithm, we observe a marked improvement in probability gain.}
\label{fig:optimal_airmass}
\end{figure}

The telescope slew rate also influences the amount of improvement the slew optimization algorithm has over the modified optimal algorithm.  Table \ref{table:slew_rate_params} indicates that slew optimization is vital for slowly slewing telescopes that are more likely only to cover a small portion of the patch within the time constraint due to time spent slewing.  Faster slewing telescopes tend to have a larger cumulative slew angle at the end of the slew-optimized schedule because they can afford to perform larger slews between tiles than slowly slewing telescopes, within the time constraint.  Table \ref{table:slew_rate_params} shows that even for telescopes with fast slew rates of 10 deg./s, slew optimization can reduce the overall time the telescope spends slewing.

\begin{table}[h!]
\begin{center}
\begin{tabular}{ccccc}
\hline\hline
$v_{slew}$ (deg./s) & algorithm & prob. & $\sigma_{slew} (deg.)$ & $t_{slew} (min.)$ \\
\hline
2 & optimal & 0.683 & 2459 & 23.42 \\
\hline
 & slewopt & 0.714 & 2095 & 17.47 \\
\hline
5 & optimal & 0.728 & 3099 & 10.33 \\
\hline
 & slewopt & 0.730 & 2501 & 8.33 \\
\hline
10 & optimal & 0.737 & 3283 & 5.47 \\
\hline
 & slewopt	& 0.739 & 2616 & 4.31 \\
\hline\hline
\end{tabular}
\end{center}
\caption{Telescope performance with varying slew rate. The table above displays the difference between the performance of the optimal and slew optimization algorithms with slew rates (v$_{slew}$) of 2, 5, and 10 deg./s in terms of cumulative probability, slew angle in degrees, and slew time (t$_{slew}$) in minutes.  The slew times here have already been factored into the tile observation times.  The sample observatory is at the location of Swope, with a one square degree field of view. With increasing slew rate, the overall slew angle increases, as does the probability, while the slew time decreases.  As expected, slew optimization is more critical for telescopes with slower slew rates.}
\label{table:slew_rate_params}
\end{table}

In a similar fashion we check how the telescope FOV influences the performance of  slew-optimization compared to modified optimal. We vary the FOV of the telescope, assuming a slew rate of 2 deg/s, and fixing the location and other parameters to that of Swope's. Table \ref{table:fov} demonstrates the relative improvement of the slew optimization algorithm over the optimal algorithm; we note that there is more improvement for smaller FOV sizes.  Therefore, the problem of slew optimization is especially relevant for smaller FOV telescopes with slow slew rates attempting to tile large GW sky error regions.

Using a different set of patch and telescope parameters, we now perform a comparison between the \awg\ and greedy algorithms, observing a marked improvement when using the \awg\ algorithm. Figure ~\ref{fig:optimal_airmass} shows the cumulative probability coverage by the greedy and \awg\ methods for a localization by the PTF telescope. Around tile 35 the optimal algorithm deviates from greedy, appearing to acquire probability more gradually at first, but eventually surpassing the greedy algorithm in tiles and probability covered. The greedy method covers 32.8\% and \awg\ method covers 36.5\% probability. We do not account for airmass in the greedy method.  When comparing Figure~\ref{fig:optimal_airmass} to the cumulative probability acquired by the original optimal algorithm, we observe that the curve corresponding to the optimal algorithm has the same shape as the \awg\ algorithm, but deviates from the greedy curve earlier in tile number.  This indicates that the \awg\ algorithm prioritizes observing higher probability tiles earlier on in the observing schedule.

In order to test the performance of our \awg\ algorithm, we generate a distribution of GW sky localizations for binary neutron stars, assuming that the sources are distributed uniformly in volume.  As our study pertains to the ADE, in which we expect our detectors to have improved in sensitivity over the current state of GW detectors, the sky patches we use are distributed between areas of 15-225 deg$^2$.  Though our algorithm can be run with various different input parameters for the FOV, coordinates, altitudes, and other observatory-specific parameters, we choose the location and parameters of the GROWTH-India telescope, changing the original field of view to a 1$\degr \times 2\degr$ square field of view to determine in which cases the \awg\ algorithm shows an improvement over the greedy algorithm.  We keep the observing horizon at an altitude of $30\degr$ (airmass $\sim 2$), and the observations are performed when the sun is below $-12\degr$ altitude. We assume that the luminosity of the optical counterpart does not change with time, which fixes an exposure time of 300 seconds to see a source at a distance of 100 Mpc. The exposure time changes for different localizations based on their average distance. We ran the \awg\ and greedy methods on 900 of these 3D-localizations for the GROWTH-India optical telescope. 616 out of 900 localizations were visible from the GROWTH-India observatory; the localizations at the far south were not visible.

\begin{figure}[htb]

\includegraphics[width=\columnwidth]{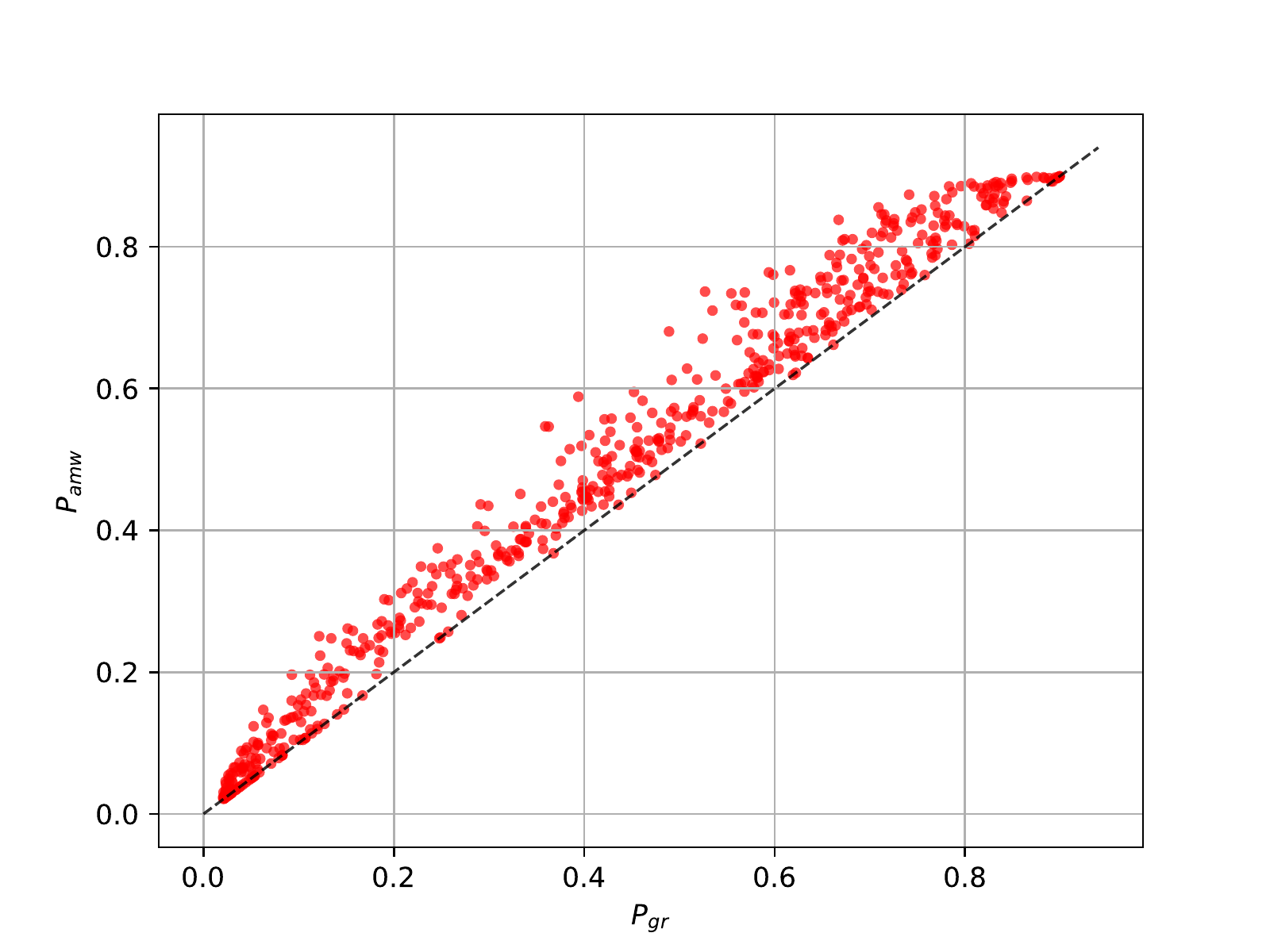}
\hspace{0.5mm}
\caption{Comparing the \awg\ method with the greedy method for 616 localizations. The x--axis is the probability covered by the greedy method, $(P_{gr})$. The y--axis is the probability covered by the \awg\ method. The dashed line represent equal coverage by both the methods.  Most of the points lie above the dashed line, indicating that P$_{amw} > P_{gr}$ for nearly all of the simulated localizations.
\label{fig:op_gr_pr}}
\end{figure}

\begin{figure}[htb]

\includegraphics[width=\columnwidth]{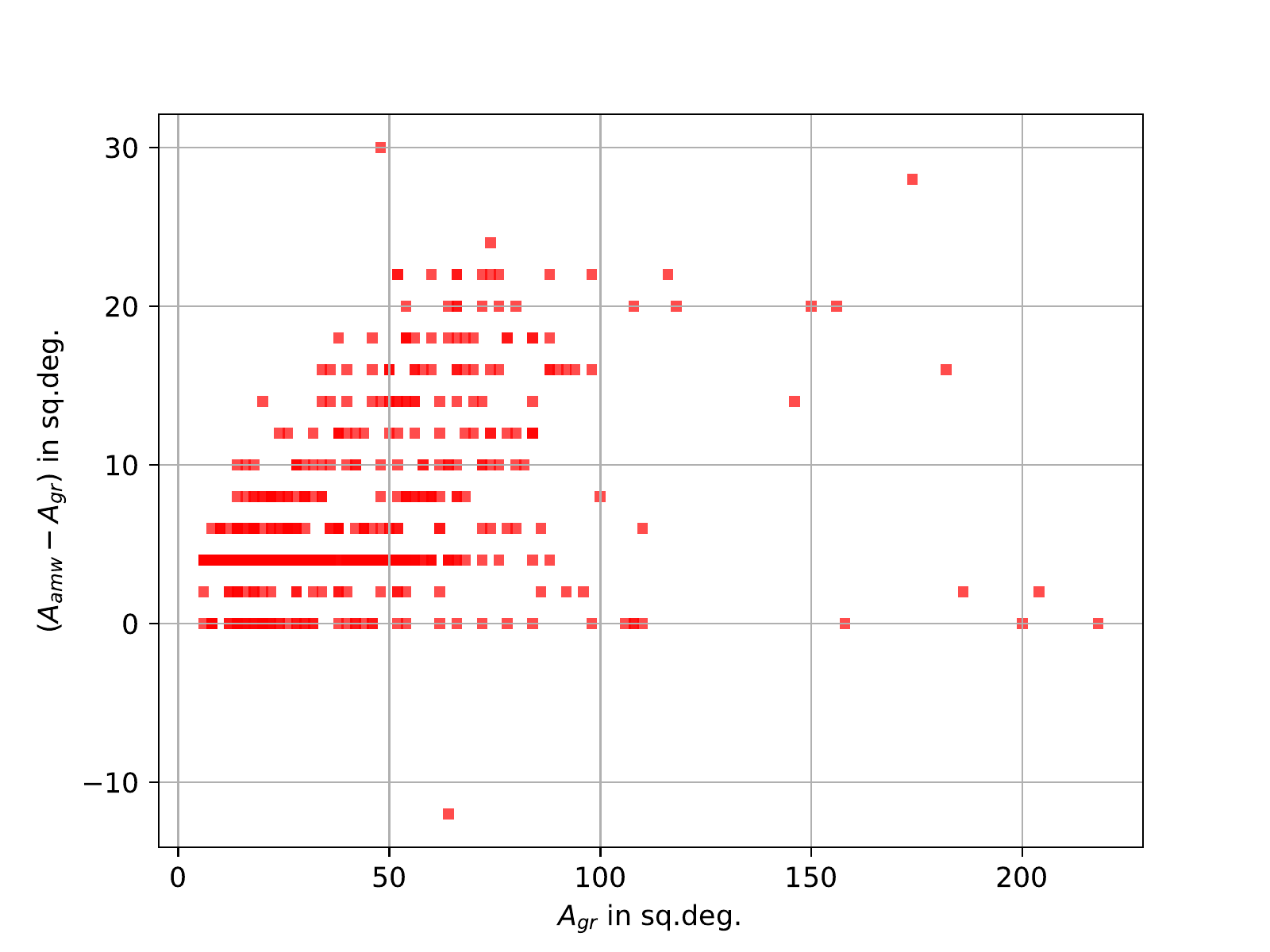}
\hspace{0.5mm}
\caption{Comparing the areal coverage between the \awg\ method and the greedy method. The x--axis is area covered by the greedy algorithm, $(A_{gr})$. The y--axis is the difference between the area covered by \awg\ algorithm and the greedy algorithm, $(A_{amw}-A_{gr})$.  Note that one data point lies below zero, indicating that in one case, the \awg\ method resulted in a loss of areal coverage.  Because the greedy method does not account for setting, or airmass, it often misses the high probability setting tiles and uses the available time to schedule low probability tiles.  While the \awg\ method always acquires more GW probability than greedy, sometimes it misses low probability tiles scheduled by greedy, resulting in a loss of areal coverage.
\label{fig:op_gr_rl_ar}}
\end{figure}

Figure~\ref{fig:op_gr_pr} presents a scatter plot of the covered total probability by the \awg\ and greedy methods. The x--axis and y--axis display the probability covered by the greedy algorithm $(P_{gr})$ and probability covered by \awg\ algorithm $(P_{amw})$ respectively. The dashed line represents equal coverage by both of the methods. Most of the points fall above the dashed line, indicating that in most cases, $P_{amw} > P_{gr}$ regardless of the amount of probability acquired by the greedy method. In Figure~\ref{fig:op_gr_rl_ar}, we compare the total areal coverage by the \awg\ method with the greedy method. Here, the x--axis is area covered by greedy algorithm, $(A_{gr})$, and the y--axis is the difference between the area covered by \awg\ algorithm and the greedy algorithm, $(A_{amw}-A_{gr})$. Each additional tile acquired by the \awg\ method over greedy adds an area of 2 square degrees because of the 2 sq. deg. field-of-view we use for these simulations.  The general trend we observe from Figures~\ref{fig:op_gr_pr} and \ref{fig:op_gr_rl_ar} is that the improvement in areal or probability coverage between the \awg\ and greedy algorithms increases with increasing P$_{gr}$ or A$_{gr}$.  This is because when patches are visible for a longer duration of time, both algorithms can cover a larger area of the patch, and the \awg\ algorithm can determine a more optimal solution to maximize the \awg\ probability.

\begin{figure}[htb]
\includegraphics[width=\columnwidth]{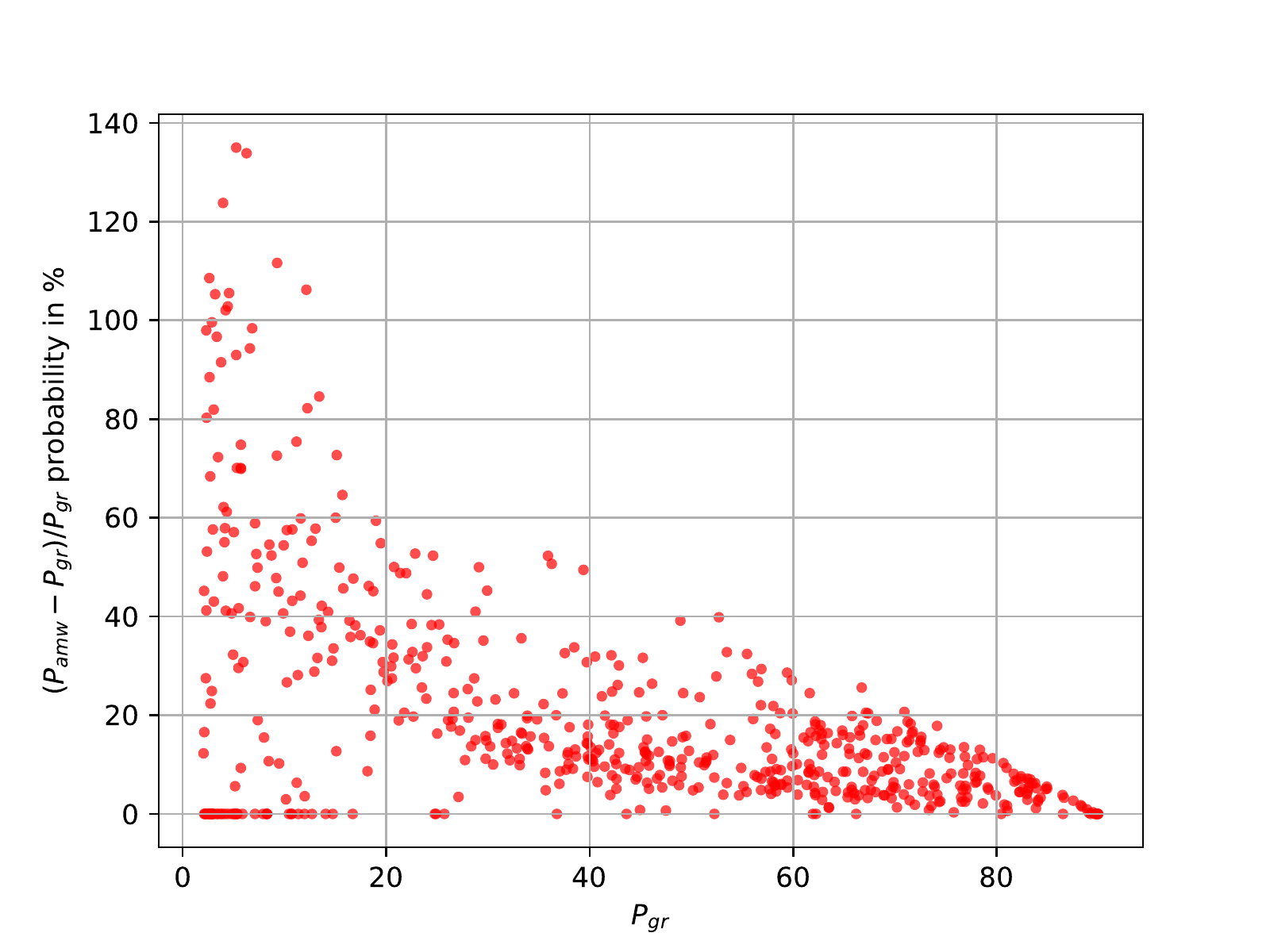}
\hspace{0.5mm}
\caption{The relative improvement in total probability coverage by the \awg\ and the greedy methods. The x--axis is the GW probability covered by greedy algorithm, $P_{gr}$, in percentage. The y--axis is the difference between the probability covered by \awg\ method and the greedy method as a function of greedy probability coverage, $(P_{amw}-P_{gr})/P_{gr}$ in percentage. Note that most of the points are lying above zero, which implies that \awg\ performs better than greedy.
\label{fig:op_gr_rl_pr_rt}}
\end{figure}

To demonstrate the relative improvement in the total probability coverage between the two methods, we show Figure~\ref{fig:op_gr_rl_pr_rt}, in which the x--axis represents the probability covered by the greedy algorithm, $P_{gr}$, while the y--axis represents the difference between the probabilities covered by the \awg\ method and the greedy method, expressed as a percentage of the greedy probability coverage  $((P_{amw}-P_{gr})/P_{gr})$. When the probability covered by the greedy method is less than $\sim 10\%$ of the total GW probability in a sky patch, then we see that the \awg\ method can cover more than double the probability acquired by the greedy method in the same patch.  However, even when the greedy method covers a large percentage ($\sim$80\%) of the GW localization, the \awg\ method is capable of covering up to 10 \% more probability than greedy, which is consistent with the earlier trend we observed.  In short, our results from simulations run to test the \awg\ method are a strong indication of the robustness of the algorithm.

\begin{table}[h!]
\begin{center}
\begin{tabular}{ccccc}
\hline\hline
FOV (deg$^2$) & algorithm & prob. & $\sigma_{slew}$ (deg.) & $t_{slew} (min.)$\\
\hline
1.0 & optimal & 0.706 & 2810 & 23.4 \\
\hline
 & slewopt & 0.714 & 2095 & 17.5 \\
\hline
2.0 & optimal & 0.888 & 4891 & 32.5 \\
\hline
 & slewopt & 0.890 & 4399 & 36.7 \\
\hline
3.0 & optimal & 0.945 & 3899 & 40.8 \\
\hline
 & slewopt &  0.949 & 3373 &  28.1 \\
\hline\hline
\end{tabular}
\end{center}
\caption{Telescope performance with varying FOV. The table above displays the difference between the performance of the optimal and slew optimization algorithms with telescope FOVs of 1.0, 2.0 and 3.0 square degrees in terms of cumulative probability, slew angle in degrees, and slew time in minutes.  The slew times here have already been factored into the tile observation times.  The sample observatory is at the location of Swope, with a slew rate of 2 deg/s. With increasing FOV, the overall slew angle increases, as does the probability, and slew time.  This is because larger FOV instruments can cover a given patch faster than a small FOV instrument, and therefore can afford to spend more of its total observing time in slewing. We find that slew optimization is more important for smaller FOV telescopes.}
\label{table:fov}
\end{table}

\section{Discussion}
\subsection{Summary}

The optimal algorithm presented in \cite{ref:rana16} provides an optimized telescope scheduling method for observing large sky-error regions that accounts for several observational constraints including the setting and rising of tiles, sun, moon and telescope time constraints. In this paper, we demonstrate the need to improve this algorithm by optimizing over slew and airmass, as the time spent slewing and additional exposure time required to resolve high airmass tiles detract from the overall time available for imaging the patch.  We present two algorithms - \awg, and slew-optimized - and describe criteria for choosing which algorithm to use to schedule observations.  One important modification we make to the original optimal algorithm is adding in an optional step in both algorithms to convolve the GW patch with a galaxy catalogue, combining the tiling and galaxy-targeted search strategies.  We demonstrate that both algorithms presented in this work are more realistic that the original optimal algorithm, and improve over the greedy algorithm not only in terms of cumulative airmass and slew, but also in overall probability acquired. 

\subsection{Caveats}

Many of the caveats of the original optimal algorithm remain true for our newly proposed algorithms, as do the procedures for overcoming these caveats.  We briefly summarize these caveats below (see \cite{ref:rana16} Sec. 4.2 for a more detailed discussion).  

We assume that the source light curve stays flat throughout the duration of the observation. GW170817 is the only BNS merger event observed so far, and its optical counterpart was first seen $\sim11$ hours after the merger. Therefore, we do not yet have any observational information about the evolution of the light curve during the 11 hours immediately following the merger. Theoretically, the optical and infrared light curves of the counterpart of the BNS merger might vary depending on the neutron star masses and equation of state. Similarly, no NSBH mergers have been observed conclusively either. Consequently, we have chosen to be agnostic about this aspect and opted for a flat light curve here. Future studies may extend this work based on any new information that arises from anticipated merger observations involving neutron stars in the coming years.

We do not account for cloudiness in our scheduling since it is outside the scope of this work.  This could be a potential weakness in most telescope scheduling algorithms -- the cloudiness constraint should be addressed in a future work.

As mentioned in the previous work, synoptic surveys will often image based on a pre-defined grid for comparison of previously imaged fields; again, here offsetting our grid tiles (placed based on the maximum GW probability) to match the pre-defined grid should not significantly affect the performance of either algorithm.

The true tile airmass, though parameterized as purely a function of the tile altitude, depends on atmospheric visibility as well as the color of filter used to conduct the observation.  We do not account for ``seeing" or filters into our airmass calculation as it would make our algorithm very telescope-specific; instead we use the formula given in Section~\ref{sec:airmass} to calculate the airmass of the tiles.

As proposed in \citep{ref:rana16}, one can adjust the final tile observation time to account for counterparts that may fade on timescales shorter than a day.  We anticipate, however, that in most cases optical observatories tiling GW or GRB localization patches will continue observation until there is confirmed non-detection.  As evidenced by the first joint GW-EM detection in 2017 \citep{ref:GW170817_MMA}, it will be difficult to determine exactly the timescale on which an optical counterpart will fade.  However, the \awg\  algorithm addresses this in part by allocating longer exposure times to tiles containing more distant galaxies.

Acknowledgements:
This work made use of the Python libraries Numpy and Matplotlib. It also made use of Astropy, a community-developed core Python package for Astronomy \citep[\url{http://www.astropy.org}]{ref:astropy}.
We would like to thank Patrick Brady, Leo Singer, Varun Bhalerao, G. C. Anupama, Om Sharan Salafia, and Shaon Ghosh for helpful discussions.We would also like to thank Michael Coughlin for carefully reading the manuscript and making useful comments in LIGO P\&P review \url{(https://dcc.ligo.org/LIGO-P1900019)}. This work was supported in part by the Navajbai Ratan Tata Grant.

\end{document}